\documentclass[submission,copyright,creativecommons]{eptcs}
\providecommand{\event}{FESCA 2017} 
\usepackage{breakurl}             
\usepackage{underscore}           
\usepackage{caption}
\usepackage{amsthm}
\usepackage{graphicx}
\usepackage{amsmath}
\usepackage{bm}
\usepackage{lineno}
\usepackage{subcaption}
\usepackage{listings}

\title{Modelling System of Systems Interface Contract Behaviour}
\author{Oldrich Faldik$^1$, Richard Payne$^2$, John Fitzgerald$^2$, and Barbora Buhnova$^3$
\institute{$^1$Faculty of Business and Economics, 
Mendel University, Brno, Czech Republic}
\institute{$^2$School of Computing Science, Newcastle University, Newcastle upon Tyne, United Kingdom}
\institute{$^3$Faculty of Informatics, Masaryk University, Brno, Czech Republic}
\email{xfaldik@mendelu.cz, richard.payne@ncl.ac.uk, john.fitzgerald@ncl.ac.uk, buhnova@fi.muni.cz}
}
\def\titlerunning{Modelling System of System Interface Contract Behaviour}
\def\authorrunning{O. Faldik, R. Payne, J. Fitzgerald \& B. Buhnova}
\begin{document}
\maketitle

\begin{abstract}
A key challenge in System of Systems (SoS) engineering is the analysis and maintenance of global properties under SoS evolution, and the integration of new constituent elements. There is a need to model the constituent systems composing a SoS in order to allow
the analysis of emergent behaviours at the SoS boundary. The Contract pattern allows the engineer to specify constrained behaviours to which constituent systems are required to conform in order to be a part of the SoS. However, the Contract pattern faces some limitations in terms of its accessibility and suitability for verifying contract compatibility.

To address these deficiencies, we propose the enrichment of the 

Contract pattern, which hitherto has been defined using SysML and the COMPASS Modelling Language (CML), by utilising SysML and  Object Constraint Language (OCL). In addition, 
we examine the potential of interface automata, a notation for improving loose coupling between interfaces of constituent systems defined according to the contract, as a means of enabling the verification of contract compatibility. The approach is demonstrated using a case study in audio/video content streaming.

\end{abstract}

\section{Introduction}
A System of Systems (SoS) is a collection of systems brought together for a task that none of the systems can accomplish on its own. Each constituent system (CS) keeps its own management, goals, and resources while coordinating within the SoS and adapting to meet SoS goals~\cite{ISO15288}.
The independence of CSs within an SoS places challenges the description of the SoS architecture, and the verification of global behaviour. 

\begin{figure}[ht!]
\centering
\includegraphics[width=0.5\textwidth]{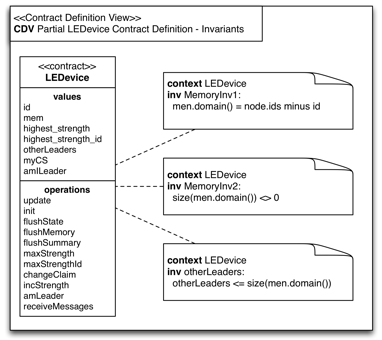}
\caption{An example Contract Definition View with OCL Invariants}
\label{OCLinvariants}
\end{figure}

There are several efforts to define architectural patterns to assist in the systematic description of SoS architectures. One such pattern is the 'Contract' pattern for specifying interfaces between CSs. A \emph{contract} constrains the behaviour, in terms of operations and their ordering, in which a CS may engage as a 'good citizen' in an SoS. For an example of a contract see Figure~\ref{OCLinvariants} where one can find contract defining operations, values and invariants for the Leader Election Device discussed later in this paper. 
The composition of such contracts can be used to verify the behaviour of the SoS as a whole. 

The Contract pattern is at the moment being realised using the SysML and the COMPASS Modelling Language (CML)~\cite{semiformal}. The SoS architectural structure is given in SysML, augmented by contractual expressions given in CML. The latter is a formal modelling language for SoSs, combining the state-based VDM and process-based Circus languages. CML may be used to specify preconditions, postconditions and invariants of contracts~\cite{maintainingesos}. SoS decriptions in SysML, with expression definitions in CML, may be completely converted to CML, permitting access to SoS analysis tools, such as simulation, model checking and theorem proving~\cite{sose}.

The previous work on the Contract Pattern has relied on the combination of SysML and CML notations. SysML is readily accessed by a range of engineering stakeholders, while the 'pure' CML is intended for the specialist SoS engineer who knows the formal notation~\cite{modelbased}. However, this approach suffers from two potential limitations. First, the use of CML may limit the ease of adoption of this approach by communities more familiar with model-based systems engineering approaches using SysML. Second, in order to complete the representation of SysML in CML, it is necessary to verify the compatibility of the contracts offered by interacting CSs; the current tools do not allow this to be done statically, and so approaches are limited to simulation (with a lack of exhaustivity) or model checking (limited by the CML  model checker)~\cite{sose}. 

\textbf{Paper Contribution:} In this paper, we examine two approaches to the potential limitations in the use of the Contract pattern outlined above. First, we consider the potential of replacing CML by the better known Object Constraint Language (OCL) which is standardised by the Object Management Group (OMG) and is used to extend the Unified Modelling Language (UML). OCL may provide potential for wider take-up of the contractual approach, and may be seen as a more natural fit with SysML. Second, we examine the use of interface automata~\cite{iautomata} as a way of increasing the opportunities for verification of contract conformance. This formalism was originally used for verification of the compatibility in component systems, and so may also be applicable in the SoS setting. 

\textbf{Paper Structure:} We first describe the areas of related work (Section~\ref{sec:rw}) and outline our approach, which utilises a case study in audio/video systems (Section~\ref{sec:outline}). Section~\ref{sec:ocl} describes the use of OCL for contract description. Section~\ref{sec:ia} discusses the application of interface automata in verifying the constitutent system compatibility. Section~\ref{sec:case-ia} applies interface automata to the case study, providing a basis for an initial evaluation, conclusions and future work (Section~\ref{sec:conc}).

\section{Related Work}
\label{sec:rw}

In this section, we consider related work in the SoS domain, and in formal model-based methods.

\paragraph{Systems of Systems}

SoSs have had a considerable amount of research in both Europe and the US in recent years. Maier~\cite{Maier98} characterises SoSs in terms of \emph{operational} and \emph{managerial independence}, \emph{distribution}, \emph{evolution} and \emph{emergent behaviour}. Maier, and later Dahmann~\cite{Dahmann&08a}, also categorise SoSs in terms of the levels of control over the CSs -- from \emph{directed SoSs} with some level of central control, to \emph{virtual SoSs} with little to no central control. An output of the COMPASS project\footnote{\url{www.compass-research.eu}} was a detailed survey of SoS engineering concepts, model-based techniques and research directions~\cite{Nielsen&15}. 

One technique for tackling complexity and understanding the composition and connections between the CSs of an SoS is through architectural modelling. In~\cite{Payne&10}, Payne and Fitzgerald survey several architecture description languages considering their applicability for modelling SoSs. SysML~\cite{SysML15}, an extended subset of UML, has found some traction in both academia and industry for modelling \emph{systems} in terms of structure, behaviour and requirements. 

SoSs pose many challenges to system engineers. Of those, the operational independence~\cite{Maier98} of the CSs introduces complications when composing constituents to examine global behaviour. \emph{Interface Contracts}~\cite{sose} have been proposed as one method of dealing with the complexity -- through detailing a collection of contracts which the CSs must conform. Arnold et al. have considered contracts in the context of SoS~\cite{arnoldetal}, using a combination of SysML/UPDM and a new language Goal and Contract Specification Language (GCSL), which combines OCL and Linear Temporal Logic (LTL). This work considers contracts at the SoS requirment level, rather than at the interfaces of individual CSs. In their survey paper~\cite{Payne&10}, Payne and Fitzgerald evaluate architecture description languages and how contracts may be represented. The term Interface Contract has much in common with the Design by Contract~\cite{Meyer88} software engineering technique. An Interface Contract details the provided and required functionality of a CS, and dictates the state- and protocol-based requirements and guarantees of those functionalities. 

In the SoS context, there is ongoing work in architectural patterns~\cite{Ingram&15a} which proposes an initial collection of topological patterns for SoS engineers, identifying aims, properties and risks. A \emph{Contract Pattern}~\cite{sose} has been defined which allows engineers to rigorously and consistently define the interface contracts of a SoS. This type of pattern considers a lower-level of abstraction to the aforementioned architectural pattern, and is considered to be an enabling pattern~\cite{COMPASSD22.6}.

\paragraph{Formalisms}

There is a wide body of literature comparing formal methods and their industrial take-up~\cite{Woodcock&09}. In SoSs, the COMPASS Modelling Language (CML) is claimed to be the first formal language defined specifically for this domain~\cite{Woodcock&12a}. CML is based on the languages
VDM~\cite{Fitzgerald:2005:VDO:1044891}, CSP~\cite{Hoare:1985:CSP:3921} and Circus~\cite{WC02}. A CML model is a collection of process definitions, each of which encapsulates a state and operations, and interacts with its environment via synchronous communications. The Contract pattern~\cite{sose} proposes the use of CML to specify preconditions, postconditions and invariants of contract operations, as well as guards/actions on contract protocol state machines. The use of pre/postconditions is intended to allow CS interfaces to be specified in terms of ranges of permitted behaviour -- important when each CS is independently owned and managed. The use of CML in this way allows an engineer to translate a profiled SysML model to a full CML model, and enables the verification afforded to CML~\cite{semiformal}.


There has been a large body of work integrating semi-formal notations such as UML and SysML with formal languages.  In ~\cite{INTOCPSD2.1a}, the authors survey  several efforts to provide a formal semantics to SysML -- through translation to different formalisms. Included in their survey is a reference to the use of \emph{interface automata}. Samir Chouali et al. \cite{CHOUALI20103} use an extended definition of interface automata to  include pre and post conditions. A  sequence of activities is performed to verify interface compatibility.

Whilst the above works apply formal notations to the verification of SysML models, the translation to these languages, or augmentation of diagrams with formal expressions, is often not a natural fit to SysML. By contrast, OCL~\cite{Warmer:2003:OCL:861416} is a standardised language, defined by OMG, who also defined SysML. OCL allows expressions, preconditions, postconditions and other constraints to be defined \emph{directly} in terms of the SysML modelling elements.

\paragraph{Contract pattern}
The Contract pattern \cite{Champlain97thecontract} was introduced as a tool that guarantees pre-conditions and post-conditions of methods and invariants that constrain the state of objects.  It means that, originally, it was designed for reliable classes. 

The Contract pattern is a wide-spread programming approach to software designing which views the construction of software as based on contracts between clients and suppliers \cite{Meyer:1992:ADC:618974.619797}. They rely on mutual commitments and benefits made explicitly expressed by statements. It has been developed in association with object-oriented programming. It is the basis for the programming language Eiffel and it is suitable for the design of component-based and agent systems \cite{Meyer:1993:SCO:162685.162705}.

\paragraph{Summary}
The contractual approach promoted in the Contract pattern has the potential to help in the integration of CSs and the verification of global behaviour. However, the  pattern has weaknesses. The first relates to the use of CML in the expression defintion. Whilst useful for translation to  CML for analysis, expressions stated in CML are not a natural fit with SysML. Second, CML does not support the analysis of contract compatibility well: simulation is not exhaustive, and the model checking capabilities are limited.

We propose an approach that firstly adopts OCL as the expression language in contract definitions. Secondly, we consider the  use of  \textit{interface automata} for analysis of contract composition.  While a mapping of SysML to interface automata has been demonstrated previously, this was limited to pure interface definitions and did not consider the contractual approach. In the next section we outline in more detail the main three areas that make up this approach.

\section{Outline of the approach}
\label{sec:outline}

In this section, we outline the notations forming the approach. First in Section ~\ref{sec:cp}, we detail the Contract Pattern and its views. Section~\ref{sec:app-ocl} outlines OCL and its main characteristics, which we will use as an extension of the Contract Pattern instead of CML. In Section~\ref{sec:rw-ia} we describe interface automata that will be later used for verification of contracts.

\subsection{Contract pattern}
\label{sec:cp}
As described in~\cite{PATTERNS}, there are many design patterns for SoS architecture. These can be categorised as \emph{architectural patterns} and \emph{enabling patterns}. The former describe specific system architecture patterns (such as a Centralised Architecture Pattern). The latter are specific constructs of modelling elements. The combination of these specific constructs and subsequent use enables many systems engineering applications. CSs may conform to multiple contracts and each may implement its contractual obligations in any way its owners choose. The Contract pattern~\cite{sose} is an enabling pattern composed of several viewpoints (see Table \ref{table:desccontractviewpoints}) in SysML and CML.

\begin{center}
\captionof{table}{Informal description of the Contract Pattern viewpoints~\cite{sose}}
\label{table:desccontractviewpoints}
    \begin{tabular}{| p{7cm} | p{8cm}  |}
    \hline
    \textbf{Name} & \textbf{Purpose of View} \\ \hline
Contractual SoS Definition Viewpoint (CSDV) &
Identifies the contracts which comprise the Contractual SoS. \\ \hline
Contract Conformance Viewpoint (CCV) & 
Identifies the constituent systems which make up the SoS and denotes the contracts to which those constituent systems conform. Includes all the contracts identified in the \textit{CSDV}. \\ \hline
Contract Connections Viewpoint (CConnV) & 
Shows connections and interfaces between contracts of the Contractual SoS. Includes all the contracts identified in the \textit{CSDV}.  \\ \hline
Contract Definition Viewpoint  (CDV) &
Defines the operations, state variables and state invariants for a single contract identified in the \textit{CSDV}.  \\ \hline
Contract Protocol Viewpoint (CPV) &
Defines the behaviour of a contract identified in the \textit{CSDV} in terms of the ordering of messages between other members of the SoS and calls to the contract operations. \\ \hline
    \end{tabular}

\end{center}

There is also the Interface pattern which is useful for defining data and interactions between CSs, but it is unsatisfactory in modelling the internal behaviour of CSs. The purpose of the Contract pattern is to enable specification of constraints on behaviours that each CS must deliver as an element of the SoS.

A limitation of the Contract pattern is that it does not precisely define which operation should and which should not be visible. In the case of the interface constituent system implementation this can lead to taking all operations as visible or as input actions. However, such an approach impairs the loose coupling of constituent system interfaces, manifesting itself by decreased flexibility of the entire system, e.g. by difficulties in replacing particular constituent systems due to other dependences. In order to solve this problem, an additional Interface definition view diagram from the Interface pattern is used, which does not improve clarity and transparency of the design because it is not stated in this contract-associated diagram.

\subsection{Object Constraint Language (OCL)}
\label{sec:app-ocl}
OCL \cite{omg2012ocl} is a language for describing constraints on a model. OCL expressions have formal semantics, and do not produce side effects influencing a described UML model \cite{Warmer:2003:OCL:861416}.

OCL offers a compromise between a natural language description and strongly formal mathematical languages. Its typical application lies in the specification of invariants for classes and types, and definition of preconditions and postconditions. It allows the definition of elements for navigation of a SysML model, referring explicitly to specfic model elements, and their attributes. This allows OCL expressions to be checked for consistency with the underlying SysML model.

OCL is a strongly typed language that defines basic types and collections. It has a well-known, OMG standardized syntax, easing the application of the contract pattern for SoS stakeholders. Its use in this context is further described in Section~\ref{sec:ocl}. 



\subsection{Interface Automata}
\label{sec:rw-ia}
\theoremstyle{definition}
\newtheorem{mydef}{Definition}

Formal description of component-based systems using \textit{interface automata} was first introduced by Alfaro and Henzinger in 2001~\cite{iautomata}. This formal notation describes the interface of a component in a component system using \textit{interface automaton} and allows verification of the component assembly. 

Every \textit{interface automaton} is composed of input actions that are modelled by methods exposed by the component to its environment, and thus they can be called. Input actions are designated by the symbol "?". Furthermore, there are output actions. These are methods required by the component from another component in a component system. Output actions are designated by the symbol "!". There are also internal actions that describe local methods of the component. Internal actions are designated by the symbol ";".

\begin{mydef}
(Interface Automaton)

An \textit{interface automaton} $A = \langle S_A,I_A,\Sigma_A^I,\Sigma_A^O,\Sigma_A^H,\delta_A \rangle$ consists of

\begin{itemize}
\item $S_A$ is a set of states.
\item $I_A \subseteq S_A$ is a set of initial states.
\item three disjoint sets $\Sigma_A^I$, $\Sigma_A^O$ and $\Sigma_A^H$ of inputs, outputs and hidden actions.
\item $\delta_A \subseteq S_A \times \Sigma_A \times S_A$
\end{itemize}

\end{mydef}

The composition condition defines, that actions of two \textit{interface automata} $A_1$ and $A_2$ are disjoint and asynchronous, except shared input and output actions. Shared actions are synchronized when $A_1$ and $A_2$ are composed. The following definition presents the composition condition.

\begin{mydef}
(Composition)

The \textit{interface automata} $A_1$, $A_2$ are composable if

$\Sigma_{A_1}^I \cap \Sigma_{A_2}^I =\Sigma_{A_1}^O \cap \Sigma_{A_2}^O = \Sigma_{A_1}^H \cap \Sigma_{A_2} = \Sigma_{A_1} \cap \Sigma_{A_2}^H = \emptyset$

If two \textit{interface automata} $A_1$ and $A_2$ are composable then 

 $Shared(A_1,A_2) = (\Sigma_{A_1}^I \cap \Sigma_{A_2}^O) \cup (\Sigma_{A_2}^I \cap \Sigma_{A_1}^O)$

\end{mydef}

The synchronous product describes parallel execution of two \textit{interface automata}.

\begin{mydef}
(Synchronized product)

Let $A_1$, $A_2$ be two composable \textit{interface automata}. The product $A_1 \otimes A_2$ is defined by

\begin{itemize}
\item $\Sigma_{A_1  \otimes A_2}^I = (\Sigma_{A_1}^I \cup \Sigma_{A_2}^I) \backslash Shared(A_1,A_2); $
\item $\Sigma_{A_1  \otimes A_2}^O = (\Sigma_{A_1}^O \cup \Sigma_{A_2}^O) \backslash
Shared(A_1,A_2); $
\item $\Sigma_{A_1  \otimes A_2}^H = \Sigma_{A_1}^H \cup \Sigma_{A_2}^H \cup
Shared(A_1,A_2); $
\item $((s_1,s_2),a,(s'_1,s'_2)) \in  \delta_{A_1 \otimes A_2}$ if
\begin{itemize}
\item $a \notin Shared(A_1,A_2)  \wedge (s_1,a,s'_1)  \in \delta_{A_1} \wedge s_2 = s'_2$
\item $a \notin Shared(A_1,A_2)  \wedge (s_2,a,s'_2)  \in \delta_{A_2} \wedge s_1 = s'_1$
\item $a \in Shared(A_1,A_2)  \wedge (s_1,a,s'_1)  \in \delta_{A_1} \wedge (s_2,a,s'_2) \in \delta_{A_2}$
\end{itemize}

\end{itemize} 

\end{mydef}

Illegal states are states which are attainable by enabling internal actions or output actions.

\begin{mydef}
(Illegal States)

Let $A_1$, $A_2$ be two composable \textit{interface automata}. The set of illegal states $Illegal(A_1,A_2)  \subseteq S_{A_1} \times S_{A_2}$ of $A_1 \otimes A_2$ is defined by $\{ (s_1,s_2) \in S_{A_1} \times  S_{A_2} | \exists a \in Shared(A_1,A_2).(a \in \Sigma_{A_1}^O (s_1) \wedge a \notin \Sigma_{A_2}^I (s_2)) \vee (a \in \Sigma_{A_2}^O (s_2) \wedge a \notin \Sigma_{A_1}^I (s_1)  )\}$.

\end{mydef}

The following algorithm describes verification of compatibility between $A_1$ and $A_2$. The result then either confirms or disproves the compatibility of components $C_1$ and $C_2$.

\begin{enumerate}

\item verify that $A_1$ and $A_2$  are composable,
\item calculate the product $A_1 \otimes A_2$,
\item calculate the set of illegal states in $A_1 \otimes A_2$,
\item calculate the bad states in $A_1 \otimes A_2$. The bad states represent states from which the illegal state are reachable by enabling only the internal action or the output actions (one suppose the existence of a helpful environment),
\item eliminate from the automaton $A_1 \otimes A_2$, the illegal state, the bad state, and the unreachable states from the initial states,
\item if the automaton $A_1 \otimes A_2$ is empty then the \textit{interface automata} $A_1$, $A_2$ are not compatible, therefore $C_1$ and $C_2$ can not assembled correctly in any environment.
\end{enumerate}

The complexity of this approach is linear on the size of $A_1$ and $A_2$ \textit{interface automata}.

\textit{Interface automata} in the Contract Pattern approach bring the possibility of verification of compatibility of contracts regarding operation visibility and complex analysis of the communication protocol of contracts.

We build upon the work of~\cite{sose} in defining Interface Contracts using the Contract Pattern. However, as detailed in Section~\ref{sec:rw}, the use of OCL may provide a more natural vehicle for combining a formal specification notation with the semi-formal SysML. Finally, we consider the use of \textit{interface automata} to enrich these contracts and to model the abstract behaviour of CSs -- complementing the state machines previously used in Interface Contracts. The use of \textit{interface automata} provides a mechanism for verifying contract compatibility.

\subsection{The Leader Election Case Study}
\label{sec:case}

In order to demonstrate the approach detailed in this paper, we use an industry-inspired case study~\cite{maintainingesos}. The study presents an audio visual (AV) network of multiple  AV devices with a network layer allowing communication between each device. 
Each device may be developed and maintained independantly, and this network exhibits the characteristic properties of SoSs. In the study each AV device must conform to a collection of contracts, each dictating required behaviours in order to be a part of the  SoS.  

Contracts exist for  streaming  of AV data,  browsing digital content and for lower-level timing issues. This study has been previously defined with the Contract Pattern ~\cite{sose}, which identifies three contracts: \emph{Browsing Device}, \emph{Streaming Device} and \emph{LE Device}. As with that paper, we concentrate on the \emph{LE Device}, in addition to the \emph{Transport Layer} contract. In~\cite{sose} the \emph{LE Device} contract is defined using SysML diagrams and CML expressions to constrain various aspects of the contract definition.

\section{OCL as an extension of SysML in CPS development}
\label{sec:ocl}

The idea of OCL applicability in SysML is based on the fact that SysML is defined as a UML profile\footnote{A UML profile provides a mechanism for customising UML models for a particular domain or platform}. This is achieved by using stereotypes and constraints applied to specific UML model elements.

OCL appears as applicable in the diagrams of Contract Definition Viewpoint and Contract Protocol Viewpoint that are basically similar to Class Diagram and State Machine Diagram. The remaining diagrams such as Contractual SoS Definition Viewpoint, Contract Conformance Viewpoint and Contract Connections Viewpoint are rather conceptual in nature and do not reach the required level of detail of description of the system where OCL should be applicable.

\subsection{Contract Definition Viewpoint}
\label{sec:ocl-cdv}

In~\cite{sose}, the Contract Definition Viewpoint (CDV) uses CML expressions to specify contract invariants, and the pre- and postconditions of contract operations. When applying OCL invariants and pre-/postconditions, we use SysML notes attached to \emph{Contract} blocks with OCL expressions. Whilst some tools are able to analyse the conditions, we leave this for future work. 

\begin{figure}[ht!]
\centering
\includegraphics[width=0.5\textwidth]{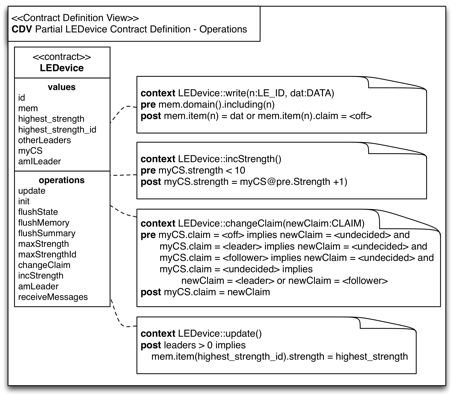}
\caption{An example Contract Definition View with OCL Operations}
\label{OCLoperations}
\end{figure}

For the purposes of clarity, we split the definition of the LE Device contract across two diagrams, shown in Figures~\ref{OCLinvariants} and~\ref{OCLoperations}. The first diagram, Figure~\ref{OCLinvariants}, 
presents the three invariants of the LE Device contract. The invariants are defined in separate SysML notes, each using the OCL expression notation for state variable invariants. The OCL statements define their \emph{context} -- here the LE Device contract -- and then a named expression. In Figure~\ref{OCLinvariants}, we see invariants which provide constraints which largely relate to the \emph{mem} variable.

Figure~\ref{OCLoperations} shows the constrained operations of the LE Device contract. It should be noted that there are several operations which do not require constraining. As above, we use SysML notes for each OCL statement -- each relating to a separate operation. The OCL statement defines the context -- here the LE Device contract, and the operation being constrained. For example, the top-most OCL statement refers to the \emph{write} operation, giving the signature and then the precondition and postcondition for the operation. These conditions are OCL expressions. 

\subsection{Contract Protocol Definition Viewpoint}
\label{sec:ocl-cpv}

\begin{figure}[ht!]
\centering
\includegraphics[width=0.8\textwidth]{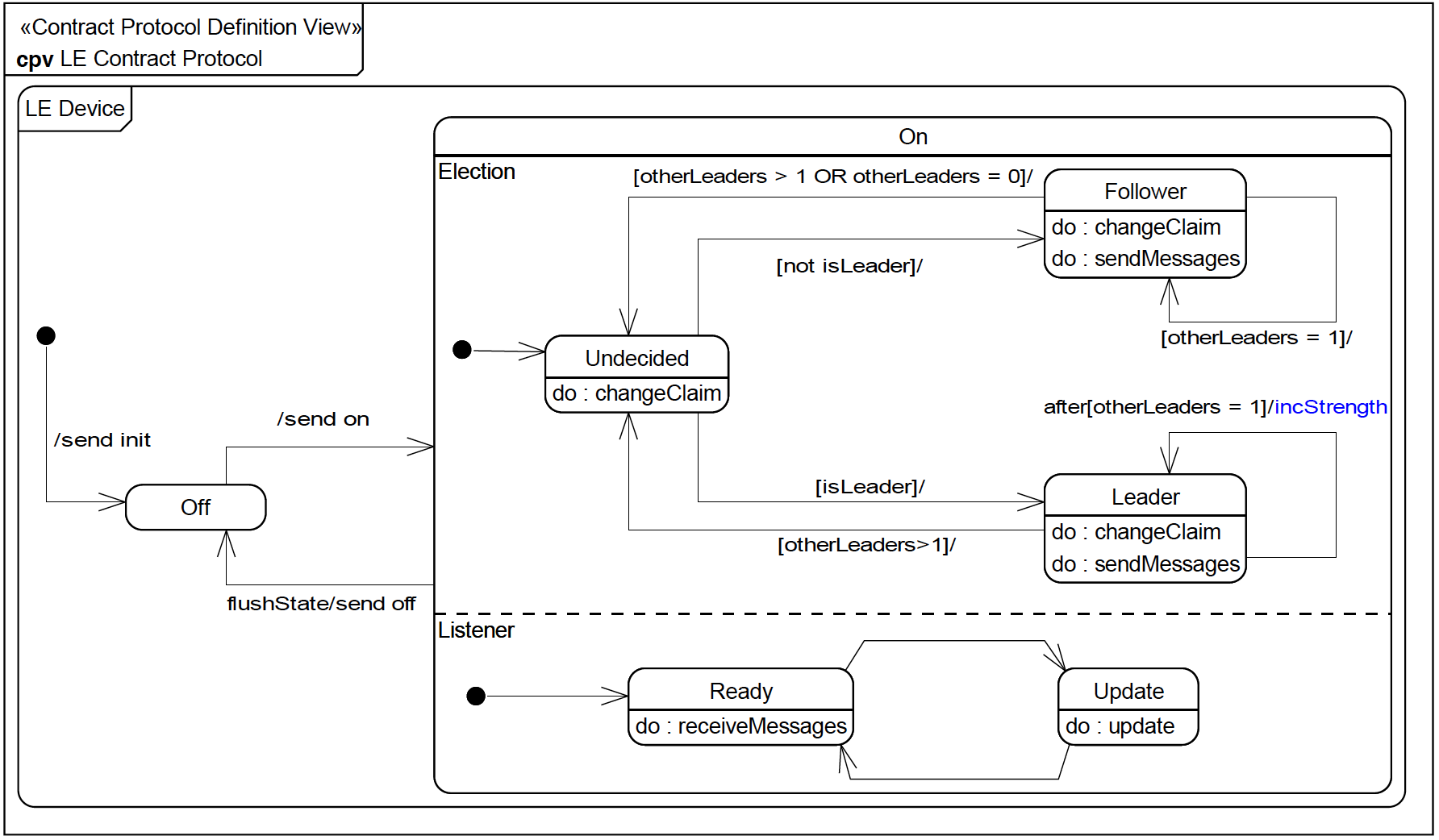}
\caption{Example Contract Protocol Definition View}
\label{OCLprotocol}
\end{figure}

Finally, we consider the Contract Protocol Definition View. Previously, in this view, CML expressions are used to define guard constraints on transitions. We propose that these guards could be defined using OCL expressions. Figure~\ref{OCLprotocol} shows the LE Device behaviour with guarded transitions. In this example, the CML and OCL expressions are the same syntactically, and therefore are unchanged.

In this section, we have demonstrated the use of OCL in the context of the Contract Pattern. We feel that the use of OCL, rather than CML provides the advantage of accessibility to the SoS engineering architecture community. 

\section{Verifying compatibility of contracts by interface automata}
\label{sec:ia}

For verification of compatibility of contracts we present an application of \textit{interface automata}, which takes into account the visibility of operations, which can exist as input, output or hidden operations. It also allows a complex view of the system using protocols for component communication. The design of \textit{interface automata} for the contract in the present study is based on the design of \textit{interface automata} described by Samir Chouali et al. \cite{CHOUALI20103} who extended the original definition by pre and post conditions, which also occur in the case of contracts. Furthermore, we extend this definition by formulation of variables, which can assume different values due to operation calls. Verification of compatibility of the two contracts is ensured by verification of their corresponding interfaces.

The following definition presents the \textit{extended interface automaton} for contract.
\begin{mydef}
(\textit{Extended interface automaton} for contract)

Let be $C$ a component of the component system $CS$, which represents a contractual specification for constituent system within a System of Systems. \textit{Extended interface automata} associated with this component is nonuple 

$A (C) = \langle S_A, I_A,\Sigma_A^I,\Sigma_A^O,\Sigma_A^H, V_A, Pre_A, Post_A, \delta_A \rangle$, which consists of

\begin{itemize}

\item $S_A$ is the set of states. 


\item $I_A \subseteq S_A$ is a set of initial states.

\item $\Sigma_A^I$, $\Sigma_A^O$ and $\Sigma_A^H$ are disjoint sets of inputs, outputs and hidden actions.

\item $V_A$ is the set of contract variables.

\item $Pre_A$	is the set of preconditions of component actions. Preconditions are specified in OCL.
\item $Post_A$ is the set of postconditions of component actions. Postconditions are specified in OCL.

\item $\delta_A$ is a plurality of transaction steps that are made on the basis of the occurrence of $a \in \Sigma$ actions in the state $s \in S$ under the valid precondition in $Pre_{A}$ and postcondition in $Post_{A}$, which might include variables from $V_{A}$. 

\end{itemize}
\end{mydef}

The following definition presents the composition of \textit{interface automata} for contract.
\begin{mydef}
(Composition of \textit{extended interface automata} for contract)

Let $A_1 = \langle S_1, I_1,\Sigma_1^I,\Sigma_1^O,\Sigma_1^H, V_1, Pre_1, Post_1, \delta_1 \rangle$ and $A_2 = \langle S_2, I_2,\Sigma_2^I,\Sigma_2^O,\Sigma_2^H, V_2, Pre_2, Post_2, \delta_2 \rangle$ be two \textit{extended interface automata}, and let's denote 
$\Sigma_{1} = \Sigma_{1}^O  \cup \Sigma_{1}^I \cup \Sigma_{1}^H$ and $\Sigma_{2} = \Sigma_{2}^O  \cup \Sigma_{2}^I \cup \Sigma_{2}^H$.

$A_1$ and $A_2$ are composable if 
$(\Sigma_{1}^I \cap \Sigma_{2}^I) =(\Sigma_{1}^O \cap \Sigma_{2}^O) = (\Sigma_{1}^H \cap \Sigma_{2}) = (\Sigma_{1} \cap \Sigma_{2}^H) = \emptyset$.

If two \textit{extended interface automata} $A_1$ and $A_2$ are composable then 

$Shared(A_1,A_2) = (\Sigma_{1}^I \cap \Sigma_{2}^O) \cup (\Sigma_{2}^I \cap \Sigma_{1}^O)$.

\end{mydef}

The following definition presents the synchronized product of two \textit{extended interface automata} for contract.
\begin{mydef}
(Synchronized product of two \textit{extended interface automata} for contract)

Let $A_1 = \langle S_1, I_1,\Sigma_1^I,\Sigma_1^O,\Sigma_1^H, V_1, Pre_1, Post_1, \delta_1 \rangle$ and $A_2 = \langle S_2, I_2,\Sigma_2^I,\Sigma_2^O,\Sigma_2^H, V_2, Pre_2, Post_2, \delta_2 \rangle$ are two composable \textit{extended interface automata}. 

The product $A_1 \otimes A_2$ is defined by $\langle S_1 \times S_2, I_1 \times I_2,\Sigma^I,\Sigma^O,\Sigma^H, V_{1} \cup V_{2}, Pre, Post, \delta \rangle$
be such that:

\begin{itemize}
\item $\Sigma^I = (\Sigma_1^I \cup \Sigma_2^I) \backslash Shared(A_1,A_2); $
\item $\Sigma^O = (\Sigma_{1}^O \cup \Sigma_{2}^O) \backslash Shared(A_1,A_2); $
\item $\Sigma^H = \Sigma_{1}^H \cup \Sigma_{2}^H \cup Shared(A_1,A_2); $
\item $((s_1,s_2),pre,a,post,(s'_1,s'_2)) \in  \delta$ if
\begin{itemize}
\item $a \notin Shared(A_1,A_2)  \wedge (s_1,pre_1,a,post_1,s'_1)  \in \delta_{1} \wedge s_2 = s'_2
 \wedge pre = pre_1 \wedge post = post_1$
\item $a \notin Shared(A_1,A_2)  \wedge (s_2,pre_2,a,post_2,s'_2)  \in \delta_{2} \wedge s_1 = s'_1
 \wedge pre = pre_2 \wedge post = post_2$
\item $a \in Shared(A_1,A_2)  \wedge ((s_1,pre_1,a,post_1,s'_1)  \in \delta_{1} \wedge a \in \Sigma^I) \wedge ((s_2,pre_2,a,post_2,s'_2)  \in \delta_{2} \wedge a \in \Sigma^O) \wedge pre = (pre_2 \wedge pre_1) \wedge post = (post_1 \wedge post_2)$
\item $a \in Shared(A_1,A_2)  \wedge ((s_1,pre_1,a,post_1,s'_1)  \in \delta_{1} \wedge a \in \Sigma^O) \wedge ((s_2,pre_2,a,post_2,s'_2)  \in \delta_{2} \wedge a \in \Sigma^I) \wedge pre = (pre_1 \wedge pre_2) \wedge post = (post_2 \wedge post_1)$
\end{itemize}
\item $Pre = Pre_{1} \cup Pre_{2} \cup \{(pre_1 \wedge pre_2)|pre_1 \in Pre_1 \wedge pre_2 \in Pre_2\}$;
\item $Post = Post_{1} \cup Post_{2} \cup \{(post_1 \wedge post_2)|post_1 \in Post_1 \wedge post_2 \in Post_2\}$.
\end{itemize} 
\end{mydef}

\medskip
The following definition presents the illegal states of two \textit{extended interface automata} for contract. A set of illegal states contains states in which shared actions between \textit{extended interface automata} are not synchronized (because required functionality by one of the automata is not provided by the other), or no transition is enabled due to the restrictions resulting from the preconditions and postconditions of the enabled transitions from the state.

\begin{mydef}
(Illegal states of two \textit{extended interface automata} for contract)

Let $A_1$, $A_2$ be two composable \textit{interface automata}. The set of illegal states $Illegal(A_1,A_2)  \subseteq S_{A_1} \times S_{A_2}$ of $A_1 \otimes A_2$ is defined by $\{ (s_1,s_2) \in S_{A_1} \times  S_{A_2} | \exists a \in Shared(A_1,A_2).(a \in \Sigma_{A_1}^O (s_1) \wedge a \notin \Sigma_{A_2}^I (s_2)) \vee (a \in \Sigma_{A_2}^O (s_2) \wedge a \notin \Sigma_{A_1}^I (s_1)  )\} \cup \{ (s_1,s_2) \in S_{A_1} \times  S_{A_2} | \forall ((s_1,s_2),pre,a,post,(s'_1,s'_2)) \in \delta_{A_1 \otimes A_2}. ((pre \equiv false) \vee (post \equiv false)) \}$.
  
\end{mydef}

\section{Translating the Leader Election Case Study to Extended Interface Automata}
\label{sec:case-ia}
In this section, we will apply the above mentioned formalism to the Leader Election study. In order to convert the contract into an \textit{extended interface automaton}, it is necessary to use the Contract Definition View which describes preconditions, values and operations (input, output, hidden), and the Contract Protocol Definition View to identify states and transitions between them. The following code presents the \textit{extended interface automata} for \textit{LE device} and \textit{Transport Layer}.\\

\centerline{\textbf{LE device}} {\scriptsize
\begin{linenumbers}
\rightlinenumbers
\setlength\linenumbersep{-40pt}
\begin{itemize} 
\item $S_{LD} = \{Off, OnFollower,OnLeader,OnUndecided,OnReady,OnUpdate\}$
\item $I_{LD} = \{Off\}$
\item $\Sigma_{LD}^I = \{receiveMessages\}$ 
\item $\Sigma_{LD}^O = \{sendMessages\}$
\item $\Sigma_{LD}^H = \{changeClaim,flushState, update,maxStrength,maxStrengthId,\\
incStrength,init,flushMemory,flushSummary,isLeader,write,turnOn,turnOff\}$
\item $V_{LD}= \{id, mem, highest\_strength, highest\_strength\_id, otherLeaders, myCS, isLeader\}$
\item $Pre_{LD} = \{LDPreCC, LDPreW, LDPreIS\}$ where: $\{   $\\
\textit{\textbf{context} LE Device::changeClaimm(newClaim : Claim)} \\
\textit{\textbf{pre LDPreCC:} $myCS.c = <off>$ $\implies$  $newc = <undecided>$} \\
\textit{and  myCS.c = $<undecided>$ $\implies$ (newc = $<leader>$ or newc = $<follower>$)}  \\
\textit{and myCS.c = $<leader>$ $\implies$ newc = $<undecided>$} \\
\textit{and myCS.c = $<follower>$ $\implies$ newc = $<undecided>$} \\\\
\textit{\textbf{context} LE Device::write(n: LE_Id, dat: DATA) \textbf{pre LDPreW:} n in set dom mem} \\
\textit{\textbf{context} LE Device::incStrength() \textbf{pre LDPreIS:} myCS.s $<$ 10} \\
$ \}$
\item $Post_{LD} = \{LDPostCC, LDPostW, LDPostIS\}$ where: $\{$  \\
\textit{\textbf{context} LE Device::changeClaimm(newClaim : Claim)  \textbf{post LDPostCC:} myCS.c = newClaim } \\
\textit{\textbf{context} LE Device::write(n: LE_Id, dat: DATA)  \textbf{post LDPostW:} mem(n) = dat or mem(n).c = $<off>$ } \\
\textit{\textbf{context} LE Device::incStrength() \textbf{post LDPostIS:} myCS.s = myCS~.s + 1 } \\
$\}$
\item $\delta_{LD} = \{$ 

\begin{itemize}
\item $Off:\bm{turnOn}:OnReady$
\item $OnReady:\bm{receiveMessages}:OnUpdate$
\item $OnUpdate:\bm{update}:OnReady$
\item $OnReady:\bm{turnOff}:Off$
\item $Off:\bm{turnOn}:OnUndecided$
\item $OnUndecided:LDPreCC:\bm{changeClaim}:LDPostCC:OnFollower$
\item $OnFollower:LDPreCC:\bm{changeClaim}:LDPostCC:OnUndecided$
\item $OnUndecided:LDPreCC:\bm{changeClaim}:LDPostCC:OnLeader$
\item $OnLeader:LDPreCC:\bm{changeClaim}:LDPostCC:OnUndecided$
\item $OnFollower:\bm{sendMessages}:OnFollower$
\item $OnLeader:\bm{sendMessages}:OnLeader$
\item $OnUndecided:\bm{turnOff}:Off$
\item $OnFollower:\bm{turnOff}:Off$
\item $OnLeader:\bm{turnOff}:Off$
\end{itemize}
$\}$ \\
\end{itemize}
\end{linenumbers}
\centerline{\textbf{Transport Layer}}
\begin{linenumbers}
\rightlinenumbers
\setlength\linenumbersep{-40pt}
\begin{itemize}
\item $S_{TL} = \{Init,Ready, CreateMessage, AddtoQueue, GetMessage, \\
CreateUnreachableMessage,TurnDeviceOn, TurnDeviceOff,SendtoDevice,ReceivedMessage\}$
\item $I_{TL} = \{Init\}$
\item $\Sigma_{TL}^I = \{sendMessages\}$ 
\item $\Sigma_{TL}^O = \{receiveMessages\}$
\item $\Sigma_{TL}^H = \{init, addToQueue, getNextMsg,createMessage, \\
AddToQueue, setDeviceOn, setDeviceOff,ready\}$
\item $V_{TL}= \{queue, devOn\}$
\item $Pre_{TL} = \{TLPreGNM, TLPreSDOF, TLPreSDON\}$ where: $\{   $\\
\textit{\textbf{context} Transport Layer::getNextMsg() \textbf{pre TLPreGNM:} queue $\rightarrow$ notEmpty} \\
\textit{\textbf{context} Transport Layer::setDeviceOff(in devId : LE_Id) \textbf{pre TLPreSDOF:} devOn[devId]$\rightarrow $notEmpty} \\ 
\textit{\textbf{context} Transport Layer::setDeviceOn(in devId : LE_Id) \textbf{pre TLPreSDON:} devOn[devId]$\rightarrow $notEmpty} \\
$ \}$
\item $Post_{TL} =  \{TLPostI, TLPostATQ\}$ where: $\{$  \\
\textit{\textbf{context} context TransportLayer::Init() \textbf{post TLPostI:} devOn.domain() = node_ids and \\
 devOn.range = {false} and queue.size() = 0} \\\\
\textit{\textbf{context} context TransportLayer::addToQueue(m:MSG) \textbf{post TLPostATQ:} \\
queue.size() = queue@pre.size() + 1 and queue.lastItem() = \\
m and queue@pre = queue(1,...,queue.size())} \\
$\}$
\item $\delta_{TL} = \{$ 

\begin{itemize}
\item $Init:\bm{init}:TLPostI:ready$
\item $Ready:\bm{sendMessages}:ReceivedMessage$
\item $ReceivedMessage:\bm{createMessage}:CreateMessage$
\item $CreateMessage:\bm{addToQueue}:TLPostATQ:AddtoQueue$ 
\item $AddtoQueue:\bm{ready}:Ready$
\item $Ready:TLPreGNM:\bm{getNextMsg}:GetMessage$ 
\item $GetMessage:\bm{receiveMessages}:SendtoDevice$
\item $SendtoDevice:\bm{ready}:Ready$
\item $GetMessage:\bm{createMessage}:CreateUnreachableMessage$
\item $SendtoDevice:\bm{createMessage}:CreateUnreachableMessage$
\item $CreateUnreachableMessage:\bm{addToQueue}:AddToQueue$
\item $AddToQueue:\bm{ready}:Ready$
\item $Ready:TLPreSDON:\bm{setDeviceOn}:TurnDeviceOn$
\item $TurnDeviceOn:\bm{ready}:Ready$
\item $Ready:TLPreSDOF:\bm{setDeviceOff}:TurnDeviceOff$
\item $TurnDeviceOff:\bm{ready}:Ready$
\end{itemize}
$\}$\\
\end{itemize}
\end{linenumbers}
}

On lines 3-6 and 43-46, the classification of operations based on whether they are input, output or hidden deserves careful attention. The application of OCL for defining precondition and postcondition are apparent from lines 8-22 and 48-60. Subsequent transitions between the states can be seen in the last sections of \textit{extended interface automata} (lines 23-38 and 61-78).  

\begin{figure}
\centering
\begin{subfigure}{.5\textwidth}
  \centering
  \includegraphics[width=1\linewidth]{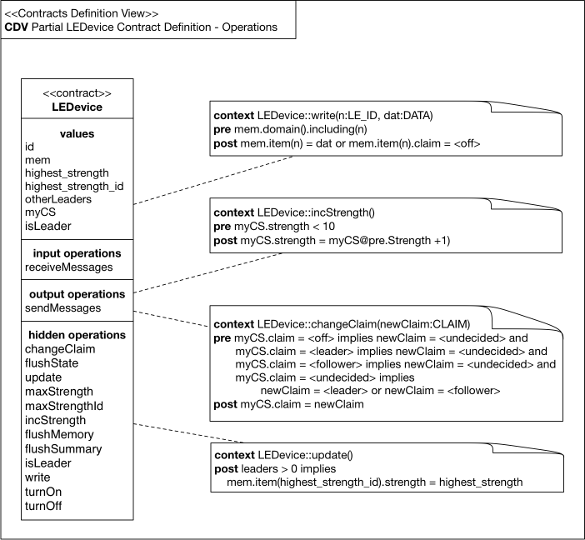}
  
  \label{fig:sub1}
\end{subfigure}%
\begin{subfigure}{.5\textwidth}
  \centering
  \includegraphics[width=0.95\linewidth]{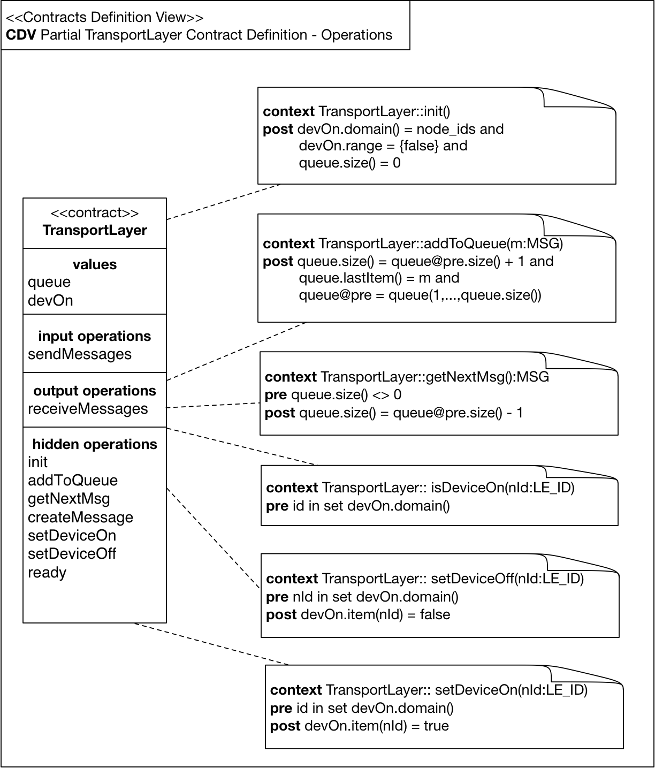}
  
  \label{fig:sub2}
\end{subfigure}
\caption{Improved Contract Definition View by separation operations}
\label{fig:improved}
\end{figure}

The classification of operations is a key element of \textit{extended interface automata}, requiring careful judgement during translation. Given the sepafration, however, we propose an improved Contract Definition View  in the Contract Pattern. As shown in Figure~\ref{fig:improved},  the Contract Definition Viewpoint identifies the input, output and hidden operations, which makes this explicit in the SysML model, and provides independence from the Interface Pattern.

In contrast to CML, during the contract translation to the interface automaton, the software engineer is led to the division of methods based on their type (input, output, hidden operations), thereby increasing the transparency. With a system described in this way we can perform classified verification operations, such as determination that constituent systems are composable (or not) by checking conditions on the actions' viability by considering their semantics. Furthermore, considering the synchronized product, we are able to determine inconsistencies between the sequences of action calls given by communicating protocols.

\section{Conclusion}
\label{sec:conc}

This paper proposes an extension of the Contract Pattern to analyse the compatibility between contracts. The paper contribution forms two main parts. Firstly, we proposed a substitution of CML for OCL language (Object Constraint Language) which is used in order to extend the SysML in the Contract Pattern. The use of OCL provides a more natural fit with the base SysML views.  In addition, we improved the Contract Pattern by separation of the input, output and hidden operations, which increased its transparency and independence on the Interface Pattern. 

Secondly, based upon the contracts defined in the Contract Pattern, we propose an improvement in interface automata. Originally designed by L. Alfaro and T. Henzinger \cite{iautomata} and modified by Samir Chouali et al. \cite{CHOUALI20103}, we adapted and extended the interface automata approach by values of the contract to represent its states. Linking CS contracts to interface automata enables the verification of contract composition. 

\paragraph{Future Work} Given the contributions of this paper, we consider several areas of future work.  Firstly, the verification of \emph{LE Device} and \emph{Transpost Layer} \textit{extended interface automata} compatibility must be demonstrated, 
based upon the set of definitions in this paper. Given manual verification, we consider automated verification. This implemention may be included as an external plug-in to the Symphony tool, and thus extended the possibilities of performing contract composition analysis in this tool platform. In addition, as mentioned in Section~\ref{sec:ocl-cdv}, the OCL expressions in this paper are defined as SysML notes. Future work would consider the conformance of CSs according to these contract specifications. Finally, we would also like to consider the application of the Contract Pattern and \textit{extended interface automata} in the specification and analysis of evolution and dynamic reconfiguration of SoSs.

\section*{Acknowledgments}
This work is supported by the INTO-CPS project funded by the European Commission’s Horizon 2020 (grant agreement 644047, http://into-cps.au.dk) and by the CPSE Labs funded by the European Union under the Smart Anything Everywhere initiative (http://www.cpse-labs.eu).

\nocite{*}
\bibliographystyle{eptcs}
\bibliography{generic}

\begin{thebibliography}{10}
\providecommand{\bibitemdeclare}[2]{}
\providecommand{\surnamestart}{}
\providecommand{\surnameend}{}
\providecommand{\urlprefix}{Available at }
\providecommand{\url}[1]{\texttt{#1}}
\providecommand{\href}[2]{\texttt{#2}}
\providecommand{\urlalt}[2]{\href{#1}{#2}}
\providecommand{\doi}[1]{doi:\urlalt{http://dx.doi.org/#1}{#1}}
\providecommand{\bibinfo}[2]{#2}

\bibitemdeclare{article}{iautomata}
\bibitem{iautomata}
\bibinfo{author}{Luca \surnamestart de~Alfaro\surnameend} \&
  \bibinfo{author}{Thomas~A. \surnamestart Henzinger\surnameend}
  (\bibinfo{year}{2001}): \emph{\bibinfo{title}{Interface Automata}}.
\newblock {\sl \bibinfo{journal}{SIGSOFT Softw. Eng. Notes}}
  \bibinfo{volume}{26}(\bibinfo{number}{5}), pp. \bibinfo{pages}{109--120},
  \doi{10.1145/503271.503226}.

\bibitemdeclare{techreport}{INTOCPSD2.1a}
\bibitem{INTOCPSD2.1a}
\bibinfo{author}{N.~\surnamestart Amalio\surnameend},
  \bibinfo{author}{R.~\surnamestart Payne\surnameend},
  \bibinfo{author}{A.~\surnamestart Cavalcanti\surnameend} \&
  \bibinfo{author}{E.~\surnamestart Brosse\surnameend} (\bibinfo{year}{2015}):
  \emph{\bibinfo{title}{{Foundations of the SysML profile for CPS modelling}}}.
\newblock \bibinfo{type}{Technical Report}, \bibinfo{institution}{{INTO-CPS}
  Deliverable, D2.1a}.

\bibitemdeclare{inproceedings}{arnoldetal}
\bibitem{arnoldetal}
\bibinfo{author}{A.~\surnamestart Arnold\surnameend},
  \bibinfo{author}{B.~\surnamestart Boyer\surnameend} \&
  \bibinfo{author}{A.~\surnamestart Legay\surnameend} (\bibinfo{year}{2013}):
  \emph{\bibinfo{title}{Contracts and Behavioral Patterns for SoS: The EU IP
  DANSE approach}}.
\newblock In \bibinfo{editor}{Kim~G. \surnamestart Larsen\surnameend},
  \bibinfo{editor}{Axel \surnamestart Legay\surnameend} \&
  \bibinfo{editor}{Ulrik \surnamestart Nyman\surnameend}, editors: {\sl
  \bibinfo{booktitle}{{\rm Proceedings 1st Workshop on} Advances in Systems of
  Systems, {\rm Rome, Italy, 16th March 2013}}}, {\sl
  \bibinfo{series}{Electronic Proceedings in Theoretical Computer Science}}
  \bibinfo{volume}{133}, \bibinfo{publisher}{Open Publishing Association}, pp.
  \bibinfo{pages}{47--66}, \doi{10.4204/EPTCS.133.6}.

\bibitemdeclare{inproceedings}{sose}
\bibitem{sose}
\bibinfo{author}{J.~\surnamestart Bryans\surnameend},
  \bibinfo{author}{J.~\surnamestart Fitzgerald\surnameend},
  \bibinfo{author}{R.~\surnamestart Payne\surnameend},
  \bibinfo{author}{A.~\surnamestart Miyazawa\surnameend} \&
  \bibinfo{author}{K.~\surnamestart Kristensen\surnameend}
  (\bibinfo{year}{2014}): \emph{\bibinfo{title}{SysML contracts for systems of
  systems}}.
\newblock In: {\sl \bibinfo{booktitle}{System of Systems Engineering (SOSE),
  2014 9th International Conference on}}, \doi{10.1109/SYSOSE.2014.6892466}.

\bibitemdeclare{inproceedings}{semiformal}
\bibitem{semiformal}
\bibinfo{author}{J.~\surnamestart Bryans\surnameend},
  \bibinfo{author}{R.~\surnamestart Payne\surnameend},
  \bibinfo{author}{J.~\surnamestart Holt\surnameend} \&
  \bibinfo{author}{S.~\surnamestart Perry\surnameend} (\bibinfo{year}{2013}):
  \emph{\bibinfo{title}{Semi-formal and formal interface specification for
  system of systems architecture}}.
\newblock In: {\sl \bibinfo{booktitle}{Systems Conference (SysCon), 2013 IEEE
  International}}, pp. \bibinfo{pages}{612--619},
  \doi{10.1109/SysCon.2013.6549946}.

\bibitemdeclare{article}{maintainingesos}
\bibitem{maintainingesos}
\bibinfo{author}{Jeremy \surnamestart Bryans\surnameend}, \bibinfo{author}{John
  \surnamestart Fitzgerald\surnameend}, \bibinfo{author}{Richard \surnamestart
  Payne\surnameend} \& \bibinfo{author}{Klaus \surnamestart
  Kristensen\surnameend} (\bibinfo{year}{2014}): \emph{\bibinfo{title}{2.2.2
  Maintaining Emergence in Systems of Systems Integration: a Contractual
  Approach using SysML}}.
\newblock {\sl \bibinfo{journal}{INCOSE International Symposium}}
  \bibinfo{volume}{24}(\bibinfo{number}{1}), pp. \bibinfo{pages}{166--181},
  \doi{10.1002/j.2334-5837.2014.tb03142.x}.

\bibitemdeclare{inproceedings}{Champlain97thecontract}
\bibitem{Champlain97thecontract}
\bibinfo{author}{Michel~De \surnamestart Champlain\surnameend}
  (\bibinfo{year}{1997}): \emph{\bibinfo{title}{The Contract Pattern}}.
\newblock In: {\sl \bibinfo{booktitle}{In Proceedings of Pattern Languages of
  Program Design 4 (PLoPD4)}}, \doi{10.1.1.38.3112}.

\bibitemdeclare{article}{CHOUALI20103}
\bibitem{CHOUALI20103}
\bibinfo{author}{S.~\surnamestart Chouali\surnameend},
  \bibinfo{author}{H.~\surnamestart Mountassir\surnameend} \&
  \bibinfo{author}{S.~\surnamestart Mouelhi\surnameend} (\bibinfo{year}{2010}):
  \emph{\bibinfo{title}{An I/O Automata-based Approach to Verify Component
  Compatibility: Application to the CyCab Car}}.
\newblock {\sl \bibinfo{journal}{Electronic Notes in Theoretical Computer
  Science}} \bibinfo{volume}{238}(\bibinfo{number}{6}), pp. \bibinfo{pages}{3
  -- 13}, \doi{10.1016/j.entcs.2010.06.002}.

\bibitemdeclare{inproceedings}{Dahmann&08a}
\bibitem{Dahmann&08a}
\bibinfo{author}{J.~S. \surnamestart Dahmann\surnameend} \&
  \bibinfo{author}{K.~J. \surnamestart Baldwin\surnameend}
  (\bibinfo{year}{2008}): \emph{\bibinfo{title}{Understanding the Current State
  of US Defense Systems of Systems and the Implications for Systems
  Engineering}}.
\newblock In: {\sl \bibinfo{booktitle}{2008 2nd Annual IEEE Systems
  Conference}}, pp. \bibinfo{pages}{1--7}, \doi{10.1109/SYSTEMS.2008.4518994}.

\bibitemdeclare{book}{Fitzgerald:2005:VDO:1044891}
\bibitem{Fitzgerald:2005:VDO:1044891}
\bibinfo{author}{J.~\surnamestart Fitzgerald\surnameend}, \bibinfo{author}{P.G.
  \surnamestart Larsen\surnameend}, \bibinfo{author}{P.~\surnamestart
  Mukherjee\surnameend}, \bibinfo{author}{N.~\surnamestart Plat\surnameend} \&
  \bibinfo{author}{M.~\surnamestart Verhoef\surnameend} (\bibinfo{year}{2005}):
  \emph{\bibinfo{title}{Validated Designs For Object-oriented Systems}}.
\newblock \bibinfo{publisher}{Springer-Verlag TELOS}, \bibinfo{address}{Santa
  Clara, CA, USA}, \doi{10.1007/b138800}.

\bibitemdeclare{book}{Hoare:1985:CSP:3921}
\bibitem{Hoare:1985:CSP:3921}
\bibinfo{author}{C.A.R. \surnamestart Hoare\surnameend} (\bibinfo{year}{1985}):
  \emph{\bibinfo{title}{Communicating Sequential Processes}}.
\newblock \bibinfo{publisher}{Prentice-Hall, Inc.}, \bibinfo{address}{Upper
  Saddle River, NJ, USA}.

\bibitemdeclare{article}{modelbased}
\bibitem{modelbased}
\bibinfo{author}{C.~\surnamestart Ingram\surnameend},
  \bibinfo{author}{R.~\surnamestart Payne\surnameend},
  \bibinfo{author}{J.~\surnamestart Fitzgerald\surnameend} \&
  \bibinfo{author}{\surnamestart L.D.Couto\surnameend} (\bibinfo{year}{2015}):
  \emph{\bibinfo{title}{Model-based Engineering of Emergence in a Collaborative
  SoS: Exploiting SysML and Formalism}}.
\newblock {\sl \bibinfo{journal}{INCOSE International Symposium}}
  \bibinfo{volume}{25}(\bibinfo{number}{1}), pp. \bibinfo{pages}{404--419},
  \doi{10.1002/j.2334-5837.2015.00071.x}.

\bibitemdeclare{article}{Ingram&15a}
\bibitem{Ingram&15a}
\bibinfo{author}{Claire \surnamestart Ingram\surnameend},
  \bibinfo{author}{Richard \surnamestart Payne\surnameend} \&
  \bibinfo{author}{John \surnamestart Fitzgerald\surnameend}
  (\bibinfo{year}{2015}): \emph{\bibinfo{title}{Architectural Modelling
  Patterns for Systems of Systems}}.
\newblock {\sl \bibinfo{journal}{INCOSE International Symposium}}
  \bibinfo{volume}{25}(\bibinfo{number}{1}), pp. \bibinfo{pages}{1177--1192},
  \doi{10.1002/j.2334-5837.2015.00123.x}.

\bibitemdeclare{manual}{ISO15288}
\bibitem{ISO15288}
\bibinfo{organization}{{ISO/IEC/IEEE}} (\bibinfo{year}{{2015}}):
  \emph{\bibinfo{title}{{15288:2015 Systems and software engineering -- System
  life cycle processes}}}.

\bibitemdeclare{article}{Maier98}
\bibitem{Maier98}
\bibinfo{author}{M.W. \surnamestart Maier\surnameend} (\bibinfo{year}{1998}):
  \emph{\bibinfo{title}{{Architecting Principles for Systems-of-Systems}}}.
\newblock {\sl \bibinfo{journal}{Systems Engineering}}
  \bibinfo{volume}{1}(\bibinfo{number}{4}), pp. \bibinfo{pages}{267--284},
  \href{http://dx.doi.org/10.1002/(SICI)1520-6858(1998)1:4<267::AID-SYS3>3.0.CO;2-D}{10.1002/(SICI)1520-6858(1998)1:4$<$267::AID-SYS3$>$3.0.CO;2-D}.

\bibitemdeclare{book}{Meyer88}
\bibitem{Meyer88}
\bibinfo{author}{B.~\surnamestart Meyer\surnameend} (\bibinfo{year}{1988}):
  \emph{\bibinfo{title}{{Object-oriented Software Construction}}}.
\newblock \bibinfo{publisher}{Prentice-Hall International}.

\bibitemdeclare{article}{Meyer:1992:ADC:618974.619797}
\bibitem{Meyer:1992:ADC:618974.619797}
\bibinfo{author}{Bertrand \surnamestart Meyer\surnameend}
  (\bibinfo{year}{1992}): \emph{\bibinfo{title}{Applying "Design by
  Contract"}}.
\newblock {\sl \bibinfo{journal}{Computer}}
  \bibinfo{volume}{25}(\bibinfo{number}{10}), pp. \bibinfo{pages}{40--51},
  \doi{10.1109/2.161279}.

\bibitemdeclare{article}{Meyer:1993:SCO:162685.162705}
\bibitem{Meyer:1993:SCO:162685.162705}
\bibinfo{author}{Bertrand \surnamestart Meyer\surnameend}
  (\bibinfo{year}{1993}): \emph{\bibinfo{title}{Systematic Concurrent
  Object-oriented Programming}}.
\newblock {\sl \bibinfo{journal}{Commun. ACM}}
  \bibinfo{volume}{36}(\bibinfo{number}{9}), pp. \bibinfo{pages}{56--80},
  \doi{10.1145/162685.162705}.

\bibitemdeclare{article}{Nielsen&15}
\bibitem{Nielsen&15}
\bibinfo{author}{C.B. \surnamestart Nielsen\surnameend}, \bibinfo{author}{P.G.
  \surnamestart Larsen\surnameend}, \bibinfo{author}{J.~\surnamestart
  Fitzgerald\surnameend}, \bibinfo{author}{J.~\surnamestart
  Woodcock\surnameend} \& \bibinfo{author}{J.~\surnamestart Peleska\surnameend}
  (\bibinfo{year}{2015}): \emph{\bibinfo{title}{Systems of Systems Engineering:
  Basic Concepts, Model-Based Techniques, and Research Directions}}.
\newblock {\sl \bibinfo{journal}{ACM Comput. Surv.}}
  \bibinfo{volume}{48}(\bibinfo{number}{2}), pp. \bibinfo{pages}{18:1--18:41},
  \doi{10.1145/2794381}.

\bibitemdeclare{}{omg2012ocl}
\bibitem{omg2012ocl}
\bibinfo{author}{\surnamestart OMG\surnameend} (\bibinfo{year}{2012}):
  \emph{\bibinfo{title}{{OMG Object Constraint Language (OCL), Version
  2.3.1}}}.
\newblock \urlprefix\url{http://www.omg.org/spec/OCL/2.3.1/}.

\bibitemdeclare{techreport}{SysML15}
\bibitem{SysML15}
\bibinfo{author}{\surnamestart OMG\surnameend} (\bibinfo{year}{2015}):
  \emph{\bibinfo{title}{{OMG Systems Modeling Language (OMG
  SysML$^{\mathrm{TM}}$)}}}.
\newblock \bibinfo{type}{Technical Report} \bibinfo{number}{Version 1.4},
  \bibinfo{institution}{Object Management Group}.
\newblock \bibinfo{note}{Http://www.omg.org/spec/SysML/1.4/}.

\bibitemdeclare{techreport}{Payne&10}
\bibitem{Payne&10}
\bibinfo{author}{R.~\surnamestart Payne\surnameend} \& \bibinfo{author}{J.S.
  \surnamestart Fitzgerald\surnameend} (\bibinfo{year}{2010}):
  \emph{\bibinfo{title}{Evaluation of architectural frameworks supporting
  contract-based specification}}.
\newblock \bibinfo{type}{Technical Report}, \bibinfo{institution}{Newcastle
  University}.

\bibitemdeclare{}{PATTERNS}
\bibitem{PATTERNS}
\bibinfo{author}{S.~\surnamestart Perry\surnameend} (\bibinfo{year}{2013}):
  \emph{\bibinfo{title}{{Report on Modelling Patterns for SoS Architectures}
  COMPASS Deliverable, D22.3, Tech. Rep.}}
\newblock \urlprefix\url{http://www.compass-research.eu/deliverables.html}.

\bibitemdeclare{techreport}{COMPASSD22.6}
\bibitem{COMPASSD22.6}
\bibinfo{author}{S.~\surnamestart Perry\surnameend},
  \bibinfo{author}{J.~\surnamestart Holt\surnameend},
  \bibinfo{author}{R.~\surnamestart Payne\surnameend},
  \bibinfo{author}{J.~\surnamestart Bryans\surnameend},
  \bibinfo{author}{C.~\surnamestart Ingram\surnameend},
  \bibinfo{author}{A.~\surnamestart Miyazawa\surnameend}, \bibinfo{author}{L.D.
  \surnamestart Couto\surnameend}, \bibinfo{author}{S.~\surnamestart
  Hallerstede\surnameend}, \bibinfo{author}{A.K. \surnamestart
  Malmos\surnameend}, \bibinfo{author}{J.~\surnamestart Iyoda\surnameend},
  \bibinfo{author}{M.~\surnamestart Cornelio\surnameend} \&
  \bibinfo{author}{J.~\surnamestart Peleska\surnameend} (\bibinfo{year}{2014}):
  \emph{\bibinfo{title}{{Final Report on SoS Architectural Models}}}.
\newblock \bibinfo{type}{Technical Report}, \bibinfo{institution}{{COMPASS}
  Deliverable, D22.6}.
\newblock \bibinfo{note}{Available at http://www.compass-research.eu/}.

\bibitemdeclare{book}{Warmer:2003:OCL:861416}
\bibitem{Warmer:2003:OCL:861416}
\bibinfo{author}{J.~\surnamestart Warmer\surnameend} \&
  \bibinfo{author}{A.~\surnamestart Kleppe\surnameend} (\bibinfo{year}{2003}):
  \emph{\bibinfo{title}{The Object Constraint Language: Getting Your Models
  Ready for MDA}}, \bibinfo{edition}{2} edition.
\newblock \bibinfo{publisher}{Addison-Wesley Longman Publishing Co., Inc.},
  \bibinfo{address}{Boston, MA, USA}.

\bibitemdeclare{inproceedings}{Woodcock&12a}
\bibitem{Woodcock&12a}
\bibinfo{author}{J.~\surnamestart Woodcock\surnameend},
  \bibinfo{author}{A.~\surnamestart Cavalcanti\surnameend},
  \bibinfo{author}{J.~\surnamestart Fitzgerald\surnameend},
  \bibinfo{author}{P.~\surnamestart Larsen\surnameend},
  \bibinfo{author}{A.~\surnamestart Miyazawa\surnameend} \&
  \bibinfo{author}{S.~\surnamestart Perry\surnameend} (\bibinfo{year}{2012}):
  \emph{\bibinfo{title}{Features of CML: A formal modelling language for
  Systems of Systems}}.
\newblock In: {\sl \bibinfo{booktitle}{2012 7th International Conference on
  System of Systems Engineering (SoSE)}}, pp. \bibinfo{pages}{1--6},
  \doi{10.1109/SYSoSE.2012.6384144}.

\bibitemdeclare{article}{Woodcock&09}
\bibitem{Woodcock&09}
\bibinfo{author}{J.~\surnamestart Woodcock\surnameend}, \bibinfo{author}{P.G.
  \surnamestart Larsen\surnameend}, \bibinfo{author}{J.~\surnamestart
  Bicarregui\surnameend} \& \bibinfo{author}{J.~\surnamestart
  Fitzgerald\surnameend} (\bibinfo{year}{2009}): \emph{\bibinfo{title}{{Formal
  Methods: Practice and Experience}}}.
\newblock {\sl \bibinfo{journal}{ACM Computing Surveys}}
  \bibinfo{volume}{41}(\bibinfo{number}{4}), pp. \bibinfo{pages}{1--36},
  \doi{10.1145/1592434.1592436}.

\bibitemdeclare{inbook}{WC02}
\bibitem{WC02}
\bibinfo{author}{Jim \surnamestart Woodcock\surnameend} \& \bibinfo{author}{Ana
  \surnamestart Cavalcanti\surnameend} (\bibinfo{year}{2002}):
  \emph{\bibinfo{title}{The Semantics of Circus}}, pp.
  \bibinfo{pages}{184--203}.
\newblock \bibinfo{publisher}{Springer Berlin Heidelberg},
  \bibinfo{address}{Berlin, Heidelberg}, \doi{10.1007/3-540-45648-1_10}.

\end{thebibliography}
\end{document}